# Transport coefficients of the fully ionized plasma with kappa-distribution and in strong magnetic field

Ran Guo[1], Jiulin Du[2]

1. *Department of Physics, College of Science, Civil Aviation University of China, Tianjin 300300, China*
2. *Department of Physics, School of Science, Tianjin University, Tianjin 300072, China*

**Abstract** Transport processes in the fully ionized plasma with kappa-distribution and in strong magnetic field are studied. By analyzing the current density and the heat flux in the $\kappa$-distributed plasma, we derive the corresponding transport coefficients, including the electric conductivity, the thermal conductivity and thermoelectric coefficient. Besides, we derive the coefficients of Hall, Nernst and Leduc-Righi effects in the $\kappa$-distributed plasma. It is shown that these new transport coefficients depend strongly on the $\kappa$-parameter and only in the limit $\kappa\to\infty$, they recover the traditional forms in the plasma based on a Maxwellian distribution. We also numerically analyze the role of the $\kappa$-parameter in the $\kappa$-dependent transport coefficients.

## 1. Introduction

In the field of space plasmas, many works about the measurements of plasma velocity distribution indicate that kappa-distributions are more suitable to model the plasma than Maxwellian velocity distributions especially at high energies in various space plasmas, for example, star's corona,[1] the solar wind,[2-4] the planetary exosphere,[5-7] the planetary magnetosphere,[8-10] and also some other observations,[11,12] etc. Therefore the kappa-distribution, as one of non-Maxwellian distributions, has already been a very interesting research topic in the related studies of astrophysical and space plasmas, such as origin of the kappa-distribution,[13-15] physical explanations of the kappa-parameter,[16,17] connections between the kappa-distribution and the nonextensive statistics,[18,19] ion-acoustic waves and dust-acoustic waves in plasmas,[20-26] and also some other plasma applications,[27-30] etc.

The velocity kappa-distribution is usually expressed as follows,[31]

$$f_\kappa(\mathbf{v}) = B_\kappa \left(1 + \frac{1}{2\kappa-3}\frac{mv^2}{kT}\right)^{-(\kappa+1)}, \qquad (1)$$

where $\kappa$ is the kappa-parameter, which satisfies $\kappa > 3/2$ in order to ensure the convergence of the second order moment for the distribution,[32] $\mathbf{v}$ is velocity vector, $m$ is mass of a particle, $k$ is Boltzmann constant, $T$ is temperature and $B_\kappa$ is the normalization constant given by

$$B_\kappa = \left[\frac{m}{(2\kappa-3)\pi kT}\right]^{\frac{3}{2}} \frac{\Gamma(\kappa+1)}{\Gamma(\kappa-\frac{1}{2})}. \qquad (2)$$

The velocity $\kappa$-distribution (1) recovers the Maxwellian velocity distribution when the $\kappa$-parameter tends to infinity. For the parameter $\kappa < \infty$, the distribution describes a non-equilibrium stationary state, and it measures the distance away from the equilibrium. In the previous works, the expression of $\kappa$-parameter is obtained in a very general physical situation which has magnetic field with the magnetic induction intensity **B**, inhomogeneous temperature $T$ and the overall bulk velocity **u** of the plasma,[16,17]

$$(\kappa - \frac{3}{2})k\nabla T = e\left(-\nabla \varphi_c + c^{-1}\mathbf{u} \times \mathbf{B}\right), \qquad (3)$$

where $e$ is the elementary charge, $\varphi_c$ is the Coulomb potential and $c$ is the speed of light. The above equation (3) suggests that the value of $\kappa$ is related not only to the temperature gradient and Coulomb potential, but also to the magnetic field by the overall bulk velocity **u**. Because the mathematical expression of kappa-distribution function has some similarity to the $q$-distribution in nonextensive statistical mechanics (NSM), it has been known that the kappa-distributed plasmas can be studied under the framework of NSM under certain conditions.[19] Therefore some properties of the $q$-distributions in NSM can also be applied to investigation of the complex plasmas.[19,33-35]

Transport is one of the most important research fields of plasmas. In transport processes, there can be many macroscopic thermodynamic "fluxes" driven by the thermodynamic "forces" produced due to the space inhomogeneity of the physical quantities in nonequilibrium plasmas. For instance, the heat flux is driven by the temperature gradient and the diffusion flux is driven by the density gradient. The transport coefficients are the linear correlation coefficients between the "fluxes" and the "forces". As we know, the transport properties of plasma depend to a large extent on the velocity distribution function of particles. Recently, the transport coefficients in the kappa-distributed plasmas have been studied in some simple physical situations, such as the Lorentz plasma and the weak ionized plasma.[27-30] In this work, we attempt to analyze the transport process of the fully ionized plasma with kappa-distribution and in strong magnetic field, and then derive the transport coefficients of these processes in the kappa-distributed plasma, which are different from those based on the Maxwellian distributed plasma and therefore are very important for understanding the new properties of transport processes in the complex plasmas with kappa-distributions.

The paper is organized as follows. In section 2, we give the basic transport equations and macroscopic laws for fully ionized plasma with a strong magnetic field. In section 3, we study the current density and the corresponding transport coefficients in the $\kappa$-distributed plasma. In section 4, we study the heat flux and the corresponding transport coefficients for electrons and ions, respectively, in the $\kappa$-distributed plasma. In section 5, we numerically show the roles of the $\kappa$-parameter in the kappa-dependent transport coefficients. Finally, in section 6 we give the conclusion. In addition, some complicated mathematical calculations in this paper are given in the Appendixes.

## 2. The transport equations and macroscopic laws

The transport equations of electrons and ions for describing the fully ionized plasma with a strong magnetic field can be expressed by

$$\frac{\partial n^{(e)}}{\partial t}+\mathbf{v}\cdot\frac{\partial n^{(e)}}{\partial \mathbf{r}}-\frac{e}{m}\left(\mathbf{E}+c^{-1}\mathbf{v}\times\mathbf{B}\right)\cdot\frac{\partial n^{(e)}}{\partial \mathbf{v}}=C_{ee}+C_{ei}, \qquad (4)$$

$$\frac{\partial n^{(i)}}{\partial t}+\mathbf{v}\cdot\frac{\partial n^{(i)}}{\partial \mathbf{r}}+\frac{ze}{M}\left(\mathbf{E}+c^{-1}\mathbf{v}\times\mathbf{B}\right)\cdot\frac{\partial n^{(i)}}{\partial \mathbf{v}}=C_{ii}+C_{ie}, \qquad (5)$$

where $n^{(e)}$ and $n^{(i)}$ are the distribution functions of electrons and ions, respectively, $m$ is electron mass, $M$ is ion mass, and $z$ is the charge number of the ion. $C_{ee}$, $C_{ei}$, $C_{ii}$, and $C_{ie}$ are, respectively, the collision terms of electron-electron, electron-ion, ion-ion and ion-electron. Here we assume electrons and ions have the same temperature. According to Onsager relation, the macroscopic laws of the plasma in electromagnetic field can be expressed [36, 37] by

$$\mathbf{E}+\frac{1}{c}\mathbf{u}\times\mathbf{B}+\frac{1}{eN_e}\nabla p_e = \frac{\mathbf{J}_\|}{\sigma_\|}+\frac{\mathbf{J}_\perp}{\sigma_\perp}+R\ \mathbf{B}\times\mathbf{J}+\alpha_\|\nabla_\| T+\alpha_\perp\nabla_\perp T+N\ \mathbf{B}\times\nabla T, \qquad (6)$$

$$\mathbf{q}+\frac{w_e}{e}\mathbf{J}=\alpha_\| T\mathbf{J}_\|+\alpha_\perp T\mathbf{J}_\perp+N\ T\mathbf{B}\times\mathbf{J}-\lambda_\|\nabla_\| T-\lambda_\perp\nabla_\perp T+L\ \mathbf{B}\times\nabla T, \qquad (7)$$

where $\mathbf{E}$ is the electric field intensity, $N_e$ is the number density of electrons, $p_e$ is the electron pressure, $\mathbf{J}$ is the electric current density and $\mathbf{q}$ is the energy flux. The subscripts $\|$ and $\perp$ denote parallel and perpendicular to the magnetic field $\mathbf{B}$. The transport coefficients, $\sigma$, $\alpha$ and $\lambda$, are the electric conductivity, the thermoelectric coefficient and the thermal conductivity, respectively. The other three coefficients, $R$, $N$ and $L$, stand for Hall, Nernst and Leduc-Righi effects, respectively.

As we know, because the coefficients of the transport parallel to $\mathbf{B}$ are the same as those in the plasma without magnetic field, in the present paper we only consider the transverse transport coefficients, the coefficients of the transport perpendicular to $\mathbf{B}$, so as to focus on the role of magnetic field in the transport coefficients. Further, in order to simplify the calculations, we assume the plasma to be in the reference frame of $\mathbf{u}=0$, which is valid because the transport coefficients under consideration in the present paper do not depend on this bulk velocity. Although this treatment may lead to the appearance of some non-inertial force terms in the kinetic equation, it has no effect on our results of the transport coefficients, which means that these non-inertial force terms can be neglected. In Eq.(6), we can take $\mathbf{E}=0$ because $\mathbf{E}$ only appear in the sum, i.e. $\mathbf{E}+(eN_e)^{-1}\nabla p_e$, and has no effect on the calculations of the transport coefficients.[36, 37] Based on the simplifications above, the macroscopic laws of the plasma can be rewritten as

$$\frac{1}{eN_e}\nabla_\perp p_e = \frac{\mathbf{J}_\perp}{\sigma_\perp}+\mathrm{R}\ \mathbf{B}\times\mathbf{J}_\perp+\alpha_\perp\nabla_\perp T+N\ \mathbf{B}\times\nabla_\perp T, \qquad (8)$$

$$\mathbf{q}+\frac{w_e}{e}\mathbf{J}_\perp = \alpha_\perp T\mathbf{J}_\perp+N\ T\mathbf{B}\times\mathbf{J}_\perp-\lambda_\perp\nabla_\perp T+L\ \mathbf{B}\times\nabla_\perp T. \qquad (9)$$

In order to obtain these transport coefficients in Eqs.(8) and (9), we need to study the current density $\mathbf{J}$ and the heat flux $\mathbf{q}$.

## 3. The electric current density and the electric conductivity

Because the ions and electrons are assumed to be the same temperature $T$, and $M \gg m$, the speed of ions is much less than that of electrons and the ions can be regarded as almost static. The contribution of ions to the current density is much less than that of electrons and so can be neglected. Therefore, the current density is given by

$$\mathbf{J} = -e \int \mathbf{v} \, d\mathbf{v} \, n^{(e)}(\mathbf{r}, \mathbf{v}, t), \tag{10}$$

where $n^{(e)}(\mathbf{r}, \mathbf{v}, t)$ is the velocity distribution function of electrons. It is assumed that the velocity distribution function of the plasma has a small disturbance $\delta n_\kappa^{(e)}$ about the $\kappa$-distribution $n_\kappa^{(e)}$ due to the strong magnetic field, namely,

$$n^{(e)}(\mathbf{r}, \mathbf{v}, t) = n_\kappa^{(e)}(\mathbf{r}, \mathbf{v}, t) + \delta n_\kappa^{(e)}(\mathbf{r}, \mathbf{v}, t), \tag{11}$$

Using (1) we have that

$$n_\kappa^{(e)}(\mathbf{r}, \mathbf{v}, t) = N_e B_\kappa \left(1 + \frac{1}{2\kappa - 3} \frac{mv^2}{kT}\right)^{-(\kappa+1)}, \tag{12}$$

where the number density $N_e$ of electrons and the temperature $T$ both can depend on the position and time. Noting that $\mathbf{J}$ equals zero for the equilibrium $\kappa$-distribution in reference frame of static plasma, so only the disturbed term gives contributions to the integral in Eq.(10),

$$\mathbf{J} = -e \int \mathbf{v} \delta n_\kappa^{(e)}(\mathbf{r}, \mathbf{v}, t) d\mathbf{v}. \tag{13}$$

In order to find $\delta n_\kappa^{(e)}(\mathbf{r}, \mathbf{v}, t)$, usually one can expand $\delta n_\kappa^{(e)}$ as a series of $1/B$ and only consider the first two terms, i.e.

$$\delta n_\kappa^{(e)} = \delta n_\kappa^{(e,1)} + \delta n_\kappa^{(e,2)}, \tag{14}$$

where $\delta n_\kappa^{(e,1)}$ and $\delta n_\kappa^{(e,2)}$ are respectively the first- and second-order small quantities with respect to the magnetic field $\mathbf{B}$, given[36, 37] by

$$\delta n_\kappa^{(e,1)} = -\frac{1}{\omega_{Be}} \mathbf{v} \cdot \mathbf{b} \times \nabla_\perp n_\kappa^{(e)}, \tag{15}$$

$$\delta n_\kappa^{(e,2)} = \frac{1}{\omega_{Be}^2} I(\mathbf{v} \cdot \nabla_\perp n_\kappa^{(e)}). \tag{16}$$

In these equations, $\omega_{Be} = eB/(mc)$ is the electronic cyclotron frequency, $\mathbf{b} \equiv \mathbf{B}/B$ is a unit vector of the magnetic field, and $I$ is the linearized collision integral. Eq.(15) and (16) were derived under the assumption $\Omega_e \gg \omega_{Be} \gg \nu_e$, where $\Omega_e$ is the Langmuir frequency of electron and $\nu_e$ is the collision frequency. In fact, Eqs.(15) and (16) contain only the parts of the odd function for $\mathbf{v}$ in $\delta n_\kappa^{(e)}$, and the parts of the even function for $\mathbf{v}$ in $\delta n_\kappa^{(e)}$ has been discarded because the even function parts of $\delta n_\kappa^{(e)}$ have no contribution to the current density (13) and the heat flux discussed in next section. The complete expansion form of $\delta n_\kappa^{(e)}$ can be found in Refs.38-40, which has the same odd function parts as the expressions given in Eqs.(15) and (16). Therefore they should give the same results in the calculations (see more details in Appendix A).

Now let us analyze the linearized collision term in Eq.(16). The linearized collision term can be written as two parts: the electron-ion term $I_{ei}$ and the electron-electron term $I_{ee}$. Here we only need to consider the electron-ion term in the second-order approximation because the electron-electron collision is momentum conservation and has no contribution to the second-order approximation, i.e. $\int \mathbf{v} I_{ee} d\mathbf{v} = 0$. The linearized integral of the electron-ion collision term is written[36, 37] as

$$I_{ei}(\mathbf{v} \cdot \nabla_\perp n_\kappa^{(e)}) = -\nu_{ei}(\mathbf{v} \cdot \nabla_\perp) n_\kappa^{(e)}, \tag{17}$$

where $\nu_{ei}(v)$ is the electron-ion collision frequency,

$$\nu_{ei}(v) = \frac{4\pi z e^4 N_e L_e}{m^2 v^3}, \tag{18}$$

with the electron-ion scattering factor $L_e \equiv \ln \Lambda$. Thus in the calculation of the current density, the second order approximation can be expressed as

$$\delta n_\kappa^{(e,2)} = \frac{1}{\omega_{Be}^2} I_{ei}(\mathbf{v} \cdot \nabla_\perp n_\kappa^{(e)}) = -\frac{1}{\omega_{Be}^2} \nu_{ei}(\mathbf{v} \cdot \nabla_\perp) n_\kappa^{(e)}. \tag{19}$$

Taking Eqs.(14)-(19) into (13) and using the following equation,

$$\int \mathbf{v}\mathbf{v} \cdot \mathbf{F}(v) d\mathbf{v} = \frac{1}{3} \int v^2 \mathbf{F}(v) d\mathbf{v},$$

we derive the expressions of the first- and second-order approximations of the current density,

$$\mathbf{J} = \mathbf{J}^{(1)} + \mathbf{J}^{(2)}, \tag{20}$$

$$\mathbf{J}^{(1)} = \frac{mc}{3B} \mathbf{b} \times \nabla_\perp \left( N_e \langle v^2 \rangle_\kappa \right) = \frac{c}{B} \mathbf{b} \times \nabla_\perp p_e, \tag{21}$$

$$\mathbf{J}^{(2)} = \frac{eN_e}{3\omega_{Be}} \nabla_\perp \langle v^2 \nu_{ei}(v) \rangle_\kappa = g_\kappa \frac{4\sqrt{2\pi} z e^5 N_e L_e}{3\omega_{Be}^2 m^{3/2}} \nabla_\perp \left( \frac{p_e}{(kT)^{3/2}} \right), \tag{22}$$

where the average is defined as $\langle F(\mathbf{v}) \rangle_\kappa = \frac{1}{N_e} \int F(\mathbf{v}) n_\kappa^{(e)} d\mathbf{v}$, and the kappa-dependent factor is given (see Appendix B) by

$$g_\kappa = \left( \kappa - \frac{3}{2} \right)^{-\frac{1}{2}} \frac{\Gamma(\kappa)}{\Gamma\left(\kappa - \frac{1}{2}\right)}. \tag{23}$$

In the derivation above, the equation of state, $p_e \equiv (1/3) m_e N_e \langle v^2 \rangle_\kappa = N_e kT$, is used. Comparing the current density (20) with the macroscopic law Eq.(8), we obtain the transport coefficients as follows: the Hall coefficient,

$$R_\kappa = -\frac{1}{N_e e c}, \tag{24}$$

the electric conductivity,

$$\sigma_{\kappa,\perp} = g_\kappa^{-1} \frac{3\sqrt{\pi} e^2 N_e}{\sqrt{2} m v_{ei}}, \tag{25}$$

the thermoelectric coefficient,

$$\alpha_{\kappa,\perp} = g_\kappa^2 \frac{k v_{ei}^2}{3\pi e \omega_{Be}^2}, \tag{26}$$

and the Nernst coefficient,

$$N_\kappa = -g_\kappa \frac{k v_{ei}}{\sqrt{2\pi} m c \omega_{Be}^2}, \tag{27}$$

where $v_{ei}$ is the electron-ion collision frequency at the thermal velocity $v = v_T = \sqrt{kT/m}$,

$$v_{ei} = \frac{4\pi z e^4 N_e L_e}{m^{1/2} (kT)^{3/2}}. \tag{28}$$

It is clear that these new transport coefficients depend on the $\kappa$-parameter, and when we take $\kappa \to \infty$, they all recover their traditional forms in the scene of Maxwellian distribution.

## 4. The heat flux and the thermal conductivity

The heat flux $\mathbf{q}$ in the plasma can be written as $\mathbf{q} = \mathbf{q}_e + \mathbf{q}_i$, where $\mathbf{q}_e$ is the flux due to the electrons and $\mathbf{q}_i$ is the flux due to the ions.

### 4.1 *Electronic heat flux and thermal conductivity*

We first calculate the flux due to the electrons $\mathbf{q}_e$, which is defined [36, 37] as

$$\mathbf{q}_e = \frac{m}{2} \int v^2 \mathbf{v} \delta n_\kappa^{(e)} d\mathbf{v}, \tag{29}$$

in the reference frame of $\mathbf{u} = 0$. The first-order approximation of $\mathbf{q}_e$ can be calculated directly by taking Eq.(15) into Eq.(29), i.e.

$$\mathbf{q}_e^{(1)} = \frac{m}{2} \int v^2 \mathbf{v} \delta n_\kappa^{(e,1)} d\mathbf{v}$$

$$= -\frac{m}{6\omega_{Be}} \mathbf{b} \times \left( N_e \left\langle v^4 \right\rangle_\kappa \right)$$

$$= -\frac{5kTc}{2eB} \frac{2\kappa-3}{2\kappa-5} \mathbf{b} \times \nabla_\perp p_e - \frac{5ckp_e}{2eB} \frac{2\kappa-3}{2\kappa-5} \mathbf{b} \times \nabla_\perp T, \quad (\text{for } \kappa > \frac{5}{2}). \tag{30}$$

The second-order approximation of heat flux $\mathbf{q}_e$ is that

$$\mathbf{q}_e^{(2)} = \frac{m}{2} \int v^2 \mathbf{v} \delta n_\kappa^{(e,1)} d\mathbf{v}$$

$$= \frac{m}{2\omega_{Be}^2} \int v^2 \mathbf{v} \left[ I_{ei}(\mathbf{v} \cdot \nabla_\perp n_\kappa^{(e)}) + I_{ee}(\mathbf{v} \cdot \nabla_\perp n_\kappa^{(e)}) \right] d\mathbf{v}. \tag{31}$$

So it contains two parts, one is due to the electron-ion collision, $\mathbf{q}_e^{(2,ei)}$, and the other is due to electron-electron collision, $\mathbf{q}_e^{(2,ee)}$. Thus by employing Eq.(17), we have that

$$\mathbf{q}_e^{(2,ei)} = \frac{m}{2\omega_{Be}^2}\int v^2\mathbf{v} I_{ei}(\mathbf{v}\cdot\nabla_\perp n_\kappa^{(e)})d\mathbf{v}$$

$$= -\frac{m}{2\omega_{Be}^2}\int \nu_{ei}(v) v^2\mathbf{v}\left(\mathbf{v}\cdot\nabla_\perp n_\kappa^{(e)}\right)d\mathbf{v}. \tag{32}$$

Substituting the frequency $\nu_{ei}(v)$ in (18) into (32), we get that

$$\mathbf{q}_e^{(2,ei)} = -\frac{mN_e}{6\omega_{Be}^2}\nabla_\perp \left\langle v^4 \nu_{ei}(v)\right\rangle_\kappa$$

$$= -\left(\kappa-\frac{3}{2}\right)^{\frac{1}{2}}\frac{\Gamma(\kappa-1)}{\Gamma\left(\kappa-\frac{1}{2}\right)}\frac{4\sqrt{2\pi}ze^4 N_e L_e}{3\omega_{Be}^2 m^{3/2}}\nabla_\perp\left(\frac{p_e}{\sqrt{kT}}\right). \tag{33}$$

Now, using the macroscopic laws, we can calculate the thermal conductivity coming from $\mathbf{q}_e^{(1)}$ and $\mathbf{q}_e^{(2,ei)}$. Because $\lambda_\perp$ is the transverse thermal conductivity without the current, $\mathbf{J}=0$, the macroscopic laws becomes

$$\mathbf{q} = -\lambda_\perp \nabla_\perp T + L\ \mathbf{B}\times\nabla_\perp T, \tag{34}$$

$$\frac{1}{eN_e}\nabla_\perp p_e = \alpha_\perp \nabla_\perp T + N\ \mathbf{B}\times\nabla_\perp T. \tag{35}$$

Substituting the thermoelectric coefficient in (26) and the coefficient of Nernst effect in (27) into Eq.(35), we obtain the relation,

$$\frac{1}{eN_e}\nabla_\perp p_e = g_\kappa^2\frac{k\nu_{ei}^2}{3\pi e\omega_{Be}^2}\nabla_\perp T - g_\kappa\frac{k\nu_{ei}}{\sqrt{2\pi}mc\omega_{Be}^2}\mathbf{B}\times\nabla_\perp T. \tag{36}$$

On right side of Eq.(36), the first term is a second-order small quantity of $1/\omega_{Be}$ and the second term is a first-order small quantity of $1/\omega_{Be}$ considering $\omega_{Be}=eB/(mc)$, so the first term can be neglected,

$$\nabla_\perp p_e = -g_\kappa\frac{k\nu_{ei}eN_e}{\sqrt{2\pi}mc\omega_{Be}^2}\mathbf{B}\times\nabla_\perp T$$

$$= -g_\kappa\frac{k\nu_{ei}N_e}{\sqrt{2\pi}\omega_{Be}}\mathbf{b}\times\nabla_\perp T. \tag{37}$$

Using $\mathbf{b}$ to cross the both sides of the above equation, we have

$$\mathbf{b}\times\nabla_\perp p_e = g_\kappa\frac{\nu_{ei}kN_e}{\sqrt{2\pi}\omega_{Be}}\nabla_\perp T. \tag{38}$$

Using (30) and (33) we write the heat flux of electrons, having the first-order approximation and the second-order approximation of electron-ion collision, as

$$\mathbf{q}_e^{(1)} + \mathbf{q}_e^{(2,ei)} = -\frac{5kTc}{2eB}\frac{2\kappa-3}{2\kappa-5}\mathbf{b}\times\nabla_\perp p_e - \frac{5ckp_e}{2eB}\frac{2\kappa-3}{2\kappa-5}\mathbf{b}\times\nabla_\perp T$$

$$-\left(\kappa-\frac{3}{2}\right)^{\frac{1}{2}}\frac{\Gamma(\kappa-1)}{\Gamma\left(\kappa-\frac{1}{2}\right)}\frac{4\sqrt{2\pi}ze^4 N_e L_e}{3\omega_{Be}^2 m^{3/2}}\nabla_\perp\left(\frac{p_e}{\sqrt{kT}}\right).$$

Substituting Eqs. (37) and (38) into this equation, we have that

$$\mathbf{q}_e^{(1)} + \mathbf{q}_e^{(2,ei)} = -\frac{5kTc}{2eB}\frac{2\kappa-3}{2\kappa-5}g_\kappa\frac{\nu_{ei}kN_e}{\sqrt{2\pi}\omega_{Be}}\nabla_\perp T - \frac{5ckp_e}{2eB}\frac{2\kappa-3}{2\kappa-5}\mathbf{b}\times\nabla_\perp T$$

$$+\frac{\Gamma(\kappa-1)\Gamma(\kappa)}{\left[\Gamma\left(\kappa-\frac{1}{2}\right)\right]^2}\frac{4ze^4 N_e^2 L_e \nu_{ei} k}{\omega_{Be}^3 m^{3/2}(kT)^{1/2}}\mathbf{b}\times\nabla_\perp T$$

$$+\left(\kappa-\frac{3}{2}\right)^{\frac{1}{2}}\frac{\Gamma(\kappa-1)}{\Gamma\left(\kappa-\frac{1}{2}\right)}\frac{2\sqrt{2\pi}ze^4 N_e L_e kp_e}{3\omega_{Be}^2 m^{3/2}(kT)^{3/2}}\nabla_\perp T. \quad (39)$$

On the right side of (39), the third term can be neglected because it is a third-order small quantity of $1/\omega_{Be}$. Comparing Eq.(39) with Eq.(34), we find the thermal conductivity $\lambda_{\kappa,\perp e}^{(ei)}$ coming from $\mathbf{q}_e^{(1)}$ and $\mathbf{q}_e^{(2,ei)}$,

$$\lambda_{\kappa,\perp e}^{(ei)} = \left(\kappa-\frac{3}{2}\right)^{\frac{1}{2}}\frac{\Gamma(\kappa-1)}{\Gamma\left(\kappa-\frac{1}{2}\right)}\frac{13\kappa-10}{6\kappa-15}\frac{k^2 T\nu_{ei}N_e}{\sqrt{2\pi}m\omega_{Be}^2}, \quad (\text{for } \kappa>\frac{5}{2}) \quad (40)$$

and the Leduc-Righi coefficient $L_{\kappa,e}$,

$$L_{\kappa,e} = -\left(\frac{2\kappa-3}{2\kappa-5}\right)\frac{5cp_e k}{2eB^2}, \quad (\text{for } \kappa>\frac{5}{2}). \quad (41)$$

Let us continue to calculate the second-order approximation of heat flux $\mathbf{q}_e$ due to the electron-electron collision, i.e.

$$\mathbf{q}_e^{(2,ee)} = \frac{m}{2\omega_{Be}^2}\int v^2\mathbf{v}I_{ee}(\mathbf{v}\cdot\nabla_\perp n_\kappa^{(e)})d\mathbf{v}. \quad (42)$$

The linearized collision integral $I_{ee}$ is given by Landau collision term,[37]

$$I_{ee}(\delta n_\kappa^{(e)}) = -\frac{\partial}{\partial \mathbf{v}}\cdot\mathbf{S}^{(ee)}(\delta n_\kappa^{(e)}), \quad (43)$$

where

$$\mathbf{S}^{(ee)}(\delta n_\kappa^{(e)}) = \frac{2\pi e^4 L_e}{m^2}\int \frac{w^2\mathbf{U}-\mathbf{ww}}{w^3}\cdot\left[n_\kappa^{(e)}\frac{\partial \delta n_\kappa^{\prime(e)}}{\partial \mathbf{v}'} + \delta n_\kappa^{(e)}\frac{\partial n_\kappa^{\prime(e)}}{\partial \mathbf{v}'} - n_\kappa^{\prime(e)}\frac{\partial \delta n_\kappa^{(e)}}{\partial \mathbf{v}} - \delta n_\kappa^{\prime(e)}\frac{\partial n_\kappa^{(e)}}{\partial \mathbf{v}}\right]d\mathbf{v}', \quad (44)$$

where $\mathbf{U}$ is a unit tensor, $\mathbf{w}=\mathbf{v}-\mathbf{v}'$, $n_\kappa^{\prime(e)}=n_\kappa^{\prime(e)}(\mathbf{v}')$, $\delta n_\kappa^{\prime(e)}=\delta n_\kappa^{\prime(e)}(\mathbf{v}')$, $n_\kappa^{(e)} =$ $n_\kappa^{(e)}(\mathbf{v})$, and $\delta n_\kappa^{(e)} = \delta n_\kappa^{(e)}(\mathbf{v})$. Integrating (42) by part, we have that

$$\mathbf{q}_e^{(2,ee)} = \frac{m}{2\omega_{Be}^2} \int \left(v^2 \mathbf{S} + 2\mathbf{vv}\cdot\mathbf{S}\right) d\mathbf{v}. \tag{45}$$

After tedious calculations (see the Appendix C), it is obtained that

$$\mathbf{q}_e^{(2,ee)} = -\frac{8}{3}\frac{\sqrt{\pi}e^4 L_e N_e^2 k^{1/2}}{m^{3/2}\omega_{Be}^2 T^{3/2}} C_\kappa \nabla_\perp T$$

$$+\frac{2048}{27}\frac{g_\kappa v_{ei} e^4 L_e N_e^2 \sqrt{kT}}{\sqrt{2\pi}m^{3/2}\omega_{Be}^3 T}\left[\frac{\Gamma(\kappa+1)}{\Gamma(\kappa-1/2)}\right]^2 (\kappa+1)(2\kappa-3)^{1/2} H(4,1,\kappa+2)\mathbf{b}\times\nabla_\perp T, \tag{46}$$

with the abbreviation,

$$C_\kappa = \frac{8\sqrt{2}}{3\sqrt{\pi}}\left[\frac{\Gamma(\kappa+1)}{\Gamma(\kappa-1/2)}\right]^2 (\kappa+1)\left(\kappa-\frac{3}{2}\right)^{1/2} \left\{\left[8-\frac{9}{2(\kappa+2)}\right]H(2,1,\kappa+2)\right.$$

$$-\left[28+\frac{25}{6(\kappa+2)}+\frac{44(\kappa+1)}{3(\kappa+2)}\right]H(4,1,\kappa+2)+\left[4-\frac{23}{6(\kappa+2)}\right]H(2,3,\kappa+2)$$

$$-\frac{32}{9}H(6,3,\kappa+3)-\frac{40}{9}H(4,5,\kappa+3)-\frac{8}{3}H(4,1,\kappa+3)-\frac{8}{3}H(8,1,\kappa+3)$$

$$\left.-\frac{4}{3}(\kappa+1)\left[4H(6,1,\kappa+3)+12H(4,3,\kappa+3)\right]\right\}, \tag{47}$$

and the *H* function defined by

$$H(n,m,\kappa+2) = \int_0^\infty dxdy \left(1+2x+2y+\frac{2}{3}xy+x^2+y^2\right)^{-(\kappa+2)} x^{\frac{n+1}{2}} y^{\frac{m+1}{2}}. \tag{48}$$

On the right side of Eq.(46), the second term is actually a small quantity with the order of $\left(1/\omega_{Be}^3\right)$, and so can be neglected as compared with the first term. Finally the heat flux $\mathbf{q}_e^{(2,ee)}$ is

$$\mathbf{q}_e^{(2,ee)} = -\frac{8}{3}\frac{\sqrt{\pi}e^4 L_e N_e^2 k^{1/2}}{m^{3/2}\omega_{Be}^2 T^{3/2}} C_\kappa \nabla_\perp T, \tag{49}$$

and correspondingly, the thermal conductivity of this part of the heat flux is given by

$$\lambda_{\kappa,\perp e}^{(ee)} = \frac{8}{3}\frac{\sqrt{\pi}e^4 L_e N_e^2 k^{1/2}}{m^{3/2}\omega_{Be}^2 T^{3/2}} C_\kappa. \tag{50}$$

Combining (40) with (50), the total thermal conductivity of electrons is obtained,

$$\lambda_{\kappa,\perp e} = \lambda_{\kappa,\perp e}^{(ei)} + \lambda_{\kappa,\perp e}^{(ee)}$$

$$= \left[C_\kappa + \left(\kappa-\frac{3}{2}\right)^{1/2}\frac{\Gamma(\kappa-1)}{\Gamma(\kappa-1/2)}\frac{13\kappa-10}{4\kappa-10}\frac{z}{\sqrt{2}}\right]\frac{2N_e k^2 T v_{ee}}{3\sqrt{\pi}m\omega_{Be}^2}, \tag{51}$$

where $\nu_{ee}$ is the collision frequency of electron-electron at the thermal velocity $v = v_T = \sqrt{kT/m}$,

$$\nu_{ee} = \frac{4\pi e^4 N_e L_e}{m^{1/2}(kT)^{3/2}}. \tag{52}$$

In the limit of $\kappa \to \infty$, we have $C_\kappa \to 1$ (see the Appendix E) and Eq.(51) reduces to the thermal conductivity in a Maxwell distribution[36, 37] (in the reference the factor $\sqrt{2}$ is missing which can be easily verified by sum of $\lambda_{\perp e}^{(ei)}$ and $\lambda_{\perp e}^{(ee)}$),

$$\lambda_{\perp e} = \lim_{\kappa \to \infty} \lambda_{\kappa,\perp e} = \left[1 + \frac{13}{4}\frac{z}{\sqrt{2}}\right]\frac{2N_e k^2 T \nu_{ee}}{3\sqrt{\pi} m \omega_{Be}^2}. \tag{53}$$

*4.2 Ionic heat flux and thermal conductivity*

In the same way, the ionic heat flux and thermal conductivity can be calculated assuming that the ionic cyclotron frequency $\omega_{Bi}$ satisfies $\Omega_i \gg \omega_{Bi} \gg \nu_i$, where $\Omega_i$ is the Langmuir frequency of electron and $\nu_i$ is the collision frequency.[36, 37] Following the line in the textbooks, [36, 37] the first-order approximation of the ion distribution function can be given as

$$\delta n_\kappa^{(i,1)} = \frac{1}{\omega_{Bi}}\mathbf{v}\cdot\mathbf{b}\times\left(\nabla_\perp n_\kappa^{(i)} - \frac{n_\kappa^{(i)}}{p_i}\nabla_\perp p + \frac{n_\kappa^{(i)}}{cp_i}\mathbf{J}\times\mathbf{B}\right), \tag{54}$$

where $n_\kappa^{(i)}$ is the equilibrium distribution of ions, $p_i$ is the pressure of ions, and $p = p_i + p_e$ is the total pressure. The term of $\mathbf{J}\times\mathbf{B}$ on the right of Eq.(54) can be determined by submitting it in Eq.(21), the first order approximation of current density. Taking Eq.(54) in the equation below, we can derive the first-order approximation of the ionic heat flux,

$$\begin{aligned}\mathbf{q}_i^{(1)} &= \frac{1}{2}M\int v^2 \mathbf{v}\delta n_\kappa^{(i,1)}d\mathbf{v} \\ &= \frac{M}{6\omega_{Bi}}\mathbf{b}\times\left[\nabla_\perp\left(N_i\langle v^4\rangle_\kappa\right) - \frac{N_i\langle v^4\rangle_\kappa}{p_i}\nabla_\perp p_i\right] \\ &= \frac{2\kappa-3}{2\kappa-5}\frac{5p_i kc}{2zeB^2}\mathbf{B}\times\nabla_\perp T, \quad (\text{for } \kappa > \frac{5}{2}).\end{aligned} \tag{55}$$

Comparing (55) with the macroscopic laws (34), we obtain the coefficient $L_{\kappa,i}$,

$$L_{\kappa,i} = \frac{2\kappa-3}{2\kappa-5}\frac{5p_i kc}{2zeB^2} = -\frac{L_{\kappa,e}}{z^2}, \quad (\text{for } \kappa > \frac{5}{2}). \tag{56}$$

Thus the total Leduc-Righi coefficient is written as

$$L_\kappa = L_{\kappa,e} + L_{\kappa,i} = \left(\frac{2\kappa-3}{2\kappa-5}\right)\frac{5kc}{2eB^2}\left(\frac{p_i}{z} - p_e\right). \quad (\text{for } \kappa > \frac{5}{2}) \tag{57}$$

For the second-order approximation, because the ion-electron collision hardly changes the kinetic energy of ions, we can only calculate the contribution coming from ion-ion collision, neglecting the contribution coming from ion-electron collision. The calculations are the same as those for electrons in (50). Directly, by replacing the quantities for electron with the quantities for ions in the expression (50), we find that

$$\lambda_{\kappa,\perp i} = C_\kappa \frac{2N_i k^2 T \nu_{ii}}{3\sqrt{\pi} M \omega_{Bi}^2}, \qquad (58)$$

where $\nu_{ii}$ is the collision frequency of ion-ion at the thermal velocity $v = v_T = \sqrt{kT/m}$,

$$\nu_{ii} = \frac{4\pi z^2 e^4 N_i L_i}{M^{1/2} (kT)^{3/2}}. \qquad (59)$$

Finally, the total thermal conductivity of electrons and ions is the sum of (51) and (58),

$$\begin{aligned}\lambda_{\kappa,\perp} &= \lambda_{\kappa,\perp e} + \lambda_{\kappa,\perp i} \\ &= D_\kappa \frac{2zN_e k^2 T \nu_{ee}}{3\sqrt{2\pi} m \omega_{Be}^2} + C_\kappa \frac{2k^2 T}{3\sqrt{\pi}} \left( \frac{N_e \nu_{ee}}{m \omega_{Be}^2} + \frac{N_i \nu_{ii}}{M \omega_{Bi}^2} \right),\end{aligned} \qquad (60)$$

with

$$D_\kappa = \left( \kappa - \frac{3}{2} \right)^{1/2} \frac{\Gamma(\kappa-1)}{\Gamma(\kappa-1/2)} \frac{13\kappa-10}{4\kappa-10}. \qquad (61)$$

It is clear that the new thermal conductivity in the $\kappa$-distributed plasma with strong magnetic field depends strongly on the $\kappa$-parameter and when we take the limit $\kappa \to \infty$, recovers the traditional expressions in the plasma with a Maxwell distribution.

## 5. Numerical calculations and discussions

In the sections 3 and 4, it has been shown that all the transport coefficients derived for the $\kappa$-distributed plasma with strong magnetic field can be written as a $\kappa$-dependent factor to time the traditional forms of the coefficients in the plasma with a Maxwellian distribution. In order to illustrating the properties of these new transport coefficients, numerically we give the figures of the $\kappa$-dependent factor in the electric conductivity, the thermoelectric coefficient and the thermal conductivity, shown in Figs.(1)-(3) respectively.

For the electric conductivity in (25), the $\kappa$-dependent factor is $g_\kappa^{-1}$. In Fig.(1), the factor $g_\kappa^{-1}$ varies from 0 to 1 in the range $\kappa \in (3/2, \infty)$. It is shown that the electric conductivity decreases significantly when $\kappa$-parameter decreases in the range close to $\kappa = 3/2$, which means that the $\kappa$–distributed plasma almost become insulated one when the parameter $\kappa \to 3/2$. This is because when the parameter $\kappa \to 3/2$, the plasma system approaches to the so called "q-frozen state" in which the particles are almost static, [32, 19] just like a system is close to the absolute zero. The current is not generated when any electric field is applied.

Thermoelectric coefficient (26) is depends on the factor $g_\kappa^2$, so the behavior is much similar to the $\kappa$-dependent electric conductivity. The factor $g_\kappa^2$ varies from infinity to 1 in the range $\kappa \in (3/2, \infty)$. As shown in Fig.(2), the thermoelectric conductivity increases significantly as the $\kappa$-parameter decreases in the range close to $\kappa = 3/2$. The physical explanation is still that when the plasma approaches to the "$q$-frozen state", no temperature gradient can exist, because the particles become motionless even if the electric field is applied. So in this situation the thermoelectric coefficient tends to infinity and the thermoelectric effect vanishes.

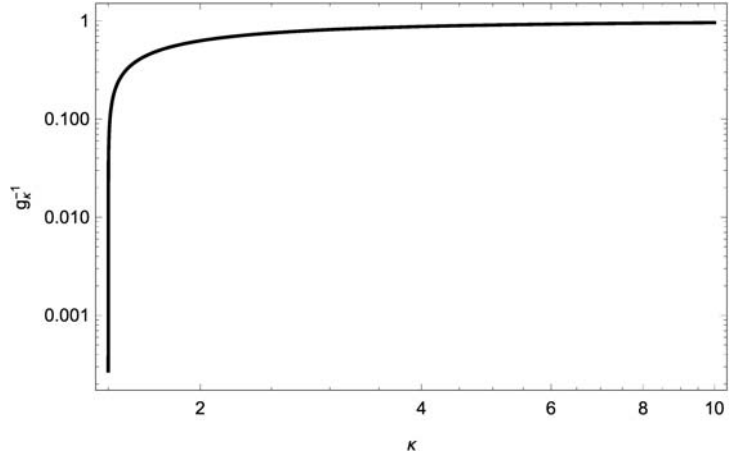

Fig.(1). Log-log figure of the kappa-modified factor $g_\kappa^{-1}$ for $\kappa \in (3/2, 10)$.

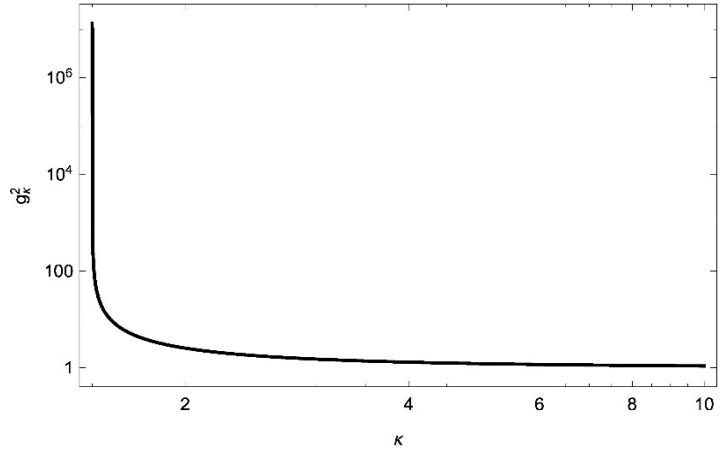

Fig.(2). Log-log figure of the kappa-modified factor $g_\kappa^2$ for $\kappa \in (3/2, 10)$.

In the thermal conductivity in (60), we see that $C_\kappa$ is the $\kappa$-dependent factor for the contribution coming from the collision between the same specific particles, while $D_\kappa$ is the $\kappa$-dependent factor for the contribution coming from the collision between the different specific particles. The property of $C_\kappa$ and $D_\kappa$ has been shown numerically in Figs. 3(a) and (b).

In Fig.(3a), the factor $C_\kappa$ decreases at first, and then increases and finally tends to 1 as the increase of the $\kappa$-parameter in the range of $\kappa \in (3/2, \infty)$. It is shown that in the range of about $\kappa < 10$, $C_\kappa$ is negative, which suggests that in that range, the gradient of temperature will generate a reverse heat flux. In Fig.(3b), we show that as increase of the $\kappa$-parameter, the factor $D_\kappa$ decreases from infinity to $13/4$ in the range of $\kappa \in (5/2, \infty)$.

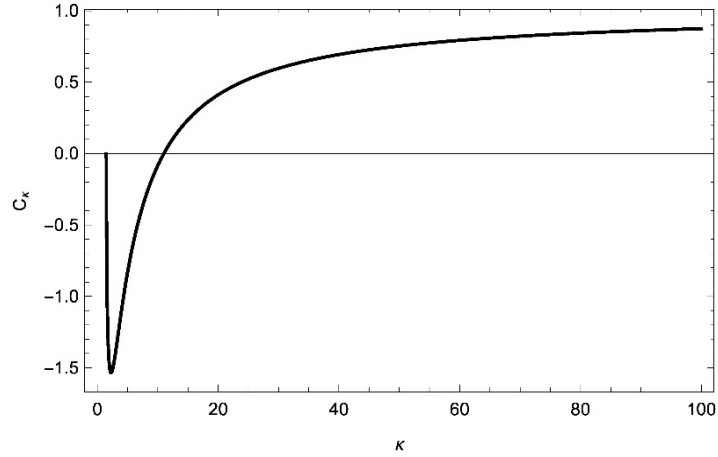

Fig.3(a). The property of $C_\kappa$ for $\kappa \in (3/2, 100)$.

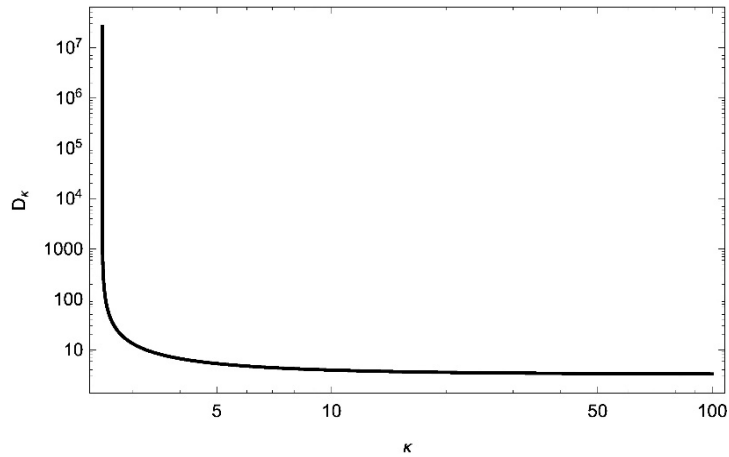

Fig.3 (b). The property of $D_\kappa$ for $\kappa \in (5/2, 100)$.

## 6. Conclusions

Transport processes are one of the most important research fields in the plasma physics and transport coefficients are key physical quantities to describe the transport processes. In this work, by using the Boltzmann equation of transport we have investigated the electric current density and the heat flux in the fully ionized and magnetized plasma with the $\kappa$-distributions. The corresponding transport coefficients are studied, mainly including the electric conductivity, the thermal conductivity and the thermoelectric coefficient. These transport coefficients for the parts perpendicular to the magnetic field are derived in the $\kappa$-distributed plasma, such as the electric conductivity $\sigma_{\kappa,\perp}$ given in (25), the thermoelectric coefficient $\alpha_{\kappa,\perp}$ given in (26), and the thermal conductivity $\lambda_{\kappa,\perp}$ given in (60). Besides, the coefficients of Hall, Nernst and Leduc-Righi effects in the $\kappa$-distributed plasma are also obtained, such as the Hall coefficient $R_\kappa$ given in Eqs.(24), the Nernst coefficient $N_\kappa$ given in (27) and the Leduc-Righi coefficient $L_\kappa$ given in (57).

We show that all the transport coefficients depend strongly on the $\kappa$-parameter in the $\kappa$-distributed plasma by the $\kappa$-dependent factors, such as $g_\kappa$ in (25)-(27), $C_\kappa$ and $D_\kappa$ in (60) etc, showing the new properties different from the transport coefficients based on the traditional statistics with a Maxwellian distribution. When we take the limit for the parameter $\kappa \to \infty$, all these coefficients recover those given in the plasma with a Maxwellian distribution.

In the Figs.(1)-(3), we have given the numerical analyses of the $\kappa$-dependent factors, $g_\kappa$, $C_\kappa$ and $D_\kappa$, clearly showing their properties which depend on the $\kappa$-parameter, so as to understand these transport processes in the $\kappa$-distributed nonequilibrium plasma when it deviates from the thermal equilibrium state.

## Acknowledgments

This work was supported by the National Natural Science Foundation of China under grant No.11775156 and was partly supported by the Special Foundation for Theoretical Physics under grant No.11747060 in the National Natural Science Foundation of China.

## Appendix A: The equivalence of the two expansions

In Refs. 38-40, the distribution of ions was derived in the first- and second-order gyroradius expansions, and the distribution of electrons was obtained by replacing the physical parameters for ions by those for electrons. These expansions derived by the drift-kinetic equation are general and complete. Here we take the distribution of electrons as an example to show that, in our calculations, these expansions are equivalent to Eq.(15) and (16) in this paper.

According to the Refs [38-40], the first two order expansions of the electron disturbance function $\delta n_\kappa^{(e)}$ can be given with respect to $1/B$ as a sum of the three parts, replacing the physical quantities for ions by those for electrons[38-40],

$$\delta n_\kappa^{(e)} = \tilde{f}^1 + \tilde{f}^2 + \tilde{f}^3, \tag{A.1}$$

with

$$\tilde{f}^1 \equiv \mathbf{v} \cdot \left[ \mathbf{g} - (\mathbf{v}_E + \mathbf{v}_M) \frac{1}{B} \frac{\partial \overline{f}}{\partial \mu} \right] + (\mathbf{v}_\perp \mathbf{v} \times \mathbf{b} + \mathbf{v} \times \mathbf{b} \mathbf{v}_\perp) : \nabla \mathbf{b} \frac{v_\parallel}{4\omega_{Be} B} \frac{\partial \overline{f}}{\partial \mu}, \tag{A.2}$$

$$\tilde{f}^2 \equiv \frac{1}{8\omega_{Be}} \mathbf{v}\mathbf{v} : \left[ \mathbf{b} \times (\mathbf{h} + \mathbf{h}^T) \cdot (\mathbf{U} + 3\mathbf{b}\mathbf{b}) - (\mathbf{U} + 3\mathbf{b}\mathbf{b}) \cdot (\mathbf{h} + \mathbf{h}^T) \times \mathbf{b} \right], \tag{A.3}$$

and

$$\tilde{f}^3 \equiv -\frac{1}{\omega_{Be}} I (\mathbf{v} \cdot \mathbf{g} \times \mathbf{b}), \tag{A.4}$$

where the denotations of the vector $\mathbf{g}$ and the dyadic tensor $\mathbf{h}$ are, respectively,

$$\mathbf{g} \equiv -\frac{1}{\omega_{Be}} \mathbf{b} \times \nabla \Big|_{\varepsilon,\mu} \overline{f} - \mathbf{v}_E \frac{\partial \overline{f}}{\partial \varepsilon}, \tag{A.5}$$

$$\mathbf{h} \equiv \nabla|_{\varepsilon,\mu} \mathbf{g} - \frac{e\mathbf{E}}{m_e} \frac{\partial \mathbf{g}}{\partial \varepsilon}. \tag{A.6}$$

In Eqs. (A.2)-(A.6), $\varepsilon = v^2/2$, $\mu = v_\perp^2/2B$ and $\nabla|_{\varepsilon,\mu}$ is the gradient operator with fixed $\varepsilon$ and $\mu$. Besides, $\mathbf{v}_E = c\mathbf{E} \times \mathbf{b}/B$ and

$$\mathbf{v}_M = -\frac{1}{\omega_{Be}} \mathbf{b} \times \left( \mu \nabla B + v_\parallel^2 \mathbf{b} \cdot \nabla \mathbf{b} + v_\parallel \partial \mathbf{b}/\partial t \right), \tag{A.7}$$

are the drift velocities. The gyrophase averaged distribution function $\bar{f}$ is defined as $\bar{f} \equiv (1/2\pi) \oint f \, d\varphi$ and its independent variables are kinetic energy $\varepsilon$, magnetic moment $\mu$, position $\mathbf{r}$ and time $t$. The superscript '$T$' denotes the transpose of a dyadic tensor.

We assume that the equilibrium electron distribution $n_\kappa^{(e)}$ depend on the $\varphi$ and $\mu$, and set the gyrophase averaged distribution function $\bar{f} = n_\kappa^{(e)}$. Therefore $\tilde{f}^1$ is simplified as

$$\tilde{f}^1 = -\frac{1}{\omega_{Be}} \mathbf{v} \cdot \mathbf{b} \times \nabla_\perp n_\kappa^{(e)} - \frac{c}{B} \mathbf{v} \cdot \mathbf{E} \times \mathbf{b} \frac{\partial n_\kappa^{(e)}}{\partial \varepsilon}, \tag{A.8}$$

with the symbol $\nabla|_{\varepsilon,\mu}$ defined [39] by

$$\nabla|_{\varepsilon,\mu} n_\kappa^{(e)} = \nabla n_\kappa^{(e)} - \nabla \mu \frac{\partial n_\kappa^{(e)}}{\partial \mu} = \nabla n_\kappa^{(e)}. \tag{A.9}$$

On the right side of Eq. (A.8), the first term is $\delta n_\kappa^{(e,1)}$ and the second term contains the electric field $\mathbf{E}$, which can be treated as $\mathbf{E}=0$ in the calculations because it appears only in the form of $\mathbf{E} + (eN_e)^{-1} \nabla p_e$. For example, in the first-order approximation $\mathbf{J}^{(1)}$ of the current density,

$$\mathbf{J}^{(1)} = -e \int \mathbf{v} \tilde{f}^1 d\mathbf{v}$$

$$= -e \int \mathbf{v} \left( \delta n_\kappa^{(e,1)} - \frac{c}{B} \mathbf{v} \cdot \mathbf{E} \times \mathbf{b} \frac{\partial n_\kappa^{(e)}}{\partial \varepsilon} \right) d\mathbf{v}$$

$$= \frac{c}{B} \mathbf{b} \times \nabla_\perp p_e + \frac{c}{B} eN_e (\mathbf{b} \times \mathbf{E})$$

$$= \frac{c}{B} \mathbf{b} \times (\nabla_\perp p_e + eN_e \mathbf{E}). \tag{A.10}$$

Therefore, in our calculation of this paper, $\tilde{f}^1$ and $\delta n_\kappa^{(e,1)}$ will give the same results.

Let us consider $\tilde{f}^2$, because only the odd function part of the distribution $\delta n_\kappa^{(e)}$ for $\mathbf{v}$ have contribution to the current density $\mathbf{J}$ and the heat flux $\mathbf{q}$, in other words, because the integrals for the even function part of $\delta n_\kappa^{(e)}$ in Eqs.(15) and (16) are all zero, we only need to consider the odd function part of $\delta n_\kappa^{(e)}$, discarding the even function part. Thus the expansion $\tilde{f}^2$ as an even function of $\mathbf{v}$ can be discarded in the current density $\mathbf{J}$ and the heat flux $\mathbf{q}$.

The expansion $\tilde{f}^3$ is the contribution coming from the linearized collisional term,

$$\mathbf{v} \cdot \mathbf{g} \times \mathbf{b} = -\frac{1}{\omega_{Be}} \mathbf{v} \cdot \left( \mathbf{b} \times \nabla n_\kappa^{(e)} \right) \times \mathbf{b} - \mathbf{v} \cdot \mathbf{v}_E \times \mathbf{b} \frac{\partial n_\kappa^{(e)}}{\partial \varepsilon}$$

$$= -\frac{1}{\omega_{Be}} \mathbf{v} \cdot \nabla_\perp n_\kappa^{(e)} + \frac{c}{B} \mathbf{v} \cdot \mathbf{E}_\perp \frac{\partial n_\kappa^{(e)}}{\partial \varepsilon}. \quad (A.11)$$

Considering that we can set $\mathbf{E}=0$ in the calculations, we have

$$\tilde{f}^3 = \frac{1}{\omega_{Be}^2} I\left( \mathbf{v} \cdot \nabla_\perp n_\kappa^{(e)} \right) = \delta n_\kappa^{(e,2)}. \quad (A.12)$$

Therefore, the gyroradius expansions $\tilde{f}^1 + \tilde{f}^2 + \tilde{f}^3$ give the same results as the expressions we used $\delta n_\kappa^{(e)} = \delta n_\kappa^{(e,1)} + \delta n_\kappa^{(e,2)}$ in our calculations of the current density $\mathbf{J}$ and the heat flux $\mathbf{q}$ with kappa-distribution function. So Eq.(15) and (16) is valid at least in the scope of this paper.

**Appendix B:** Calculation of $\langle v^n \rangle_\kappa$

$$\langle v^n \rangle_\kappa^{(e)} = B_\kappa^{(e)} \int_{-\infty}^{+\infty} v^n \left( 1 + \frac{1}{2\kappa-3} \frac{mv^2}{kT} \right)^{-(\kappa+1)} d\mathbf{v}$$

$$= B_\kappa^{(e)} \int_0^{+\infty} 4\pi v^{n+2} \left( 1 + \frac{1}{2\kappa-3} \frac{mv^2}{kT} \right)^{-(\kappa+1)} dv$$

$$= \frac{2}{\sqrt{\pi}} \left[ \frac{(2\kappa-3)kT}{m} \right]^{\frac{n}{2}} \frac{\Gamma\left(\frac{n+3}{2}\right) \Gamma\left(\kappa - \frac{n+1}{2}\right)}{\Gamma\left(\kappa - \frac{1}{2}\right)}, \quad (A.13)$$

where $n$ is a positive integer and $\kappa > (n+1)/2$ is required to ensure the convergence of the integral.

**Appendix C:** Calculation of $\mathbf{q}_{ee}^{(e,2)}$

The variable $\mathbf{v} \cdot \nabla_\perp n_\kappa$ in Eq.(42) is that

$$\mathbf{v} \cdot \nabla_\perp n_\kappa = \frac{n_\kappa}{N_e} \left( \mathbf{v} \cdot \nabla_\perp N_e \right) - \frac{3 n_\kappa}{2T} \left( \mathbf{v} \cdot \nabla_\perp T \right) + A_\kappa \frac{mv^2}{2kT^2} n_\kappa \left( \mathbf{v} \cdot \nabla_\perp T \right), \quad (A.14)$$

where

$$A_\kappa = \frac{2\kappa+2}{2\kappa-3}\left(1+\frac{1}{2\kappa-3}\frac{mv^2}{kT}\right)^{-1}. \tag{A.15}$$

Denote

$$\mathbf{s}\left(\delta n_\kappa^{(e)}\right) = n_\kappa^{(e)}\frac{\partial \delta n_\kappa'^{(e)}}{\partial \mathbf{v}'} + \delta n_\kappa^{(e)}\frac{\partial n_\kappa'^{(e)}}{\partial \mathbf{v}'} - n_\kappa'^{(e)}\frac{\partial \delta n_\kappa^{(e)}}{\partial \mathbf{v}} - \delta n_\kappa'^{(e)}\frac{\partial n_\kappa^{(e)}}{\partial \mathbf{v}}, \tag{A.16}$$

then the $\mathbf{S}^{(ee)}$ turns into

$$\mathbf{S}^{(ee)}(\delta n_\kappa^{(e)}) = \frac{2\pi e^4 L_e}{m^2}\int \frac{w^2\mathbf{U}-\mathbf{ww}}{w^3}\cdot \mathbf{s}(\delta n_\kappa^{(e)})d\mathbf{v}'. \tag{A.17}$$

The term $\mathbf{s}(\mathbf{v}\cdot\nabla_\perp n_\kappa)$ can be given directly by taking Eq.(A.14) into Eq.(A.16). After applying the variable transformation, $\mathbf{w}=\mathbf{v}-\mathbf{v}'$ and $\mathbf{G}=(\mathbf{v}+\mathbf{v}')/2$, and reductions, it turns to

$$\mathbf{s}(\mathbf{v}\cdot\nabla_\perp n_\kappa) = \left(E_{III,1}+E_{III,3}+E_{III,6}\right)(\mathbf{w}\cdot\nabla_\perp T)\mathbf{G} + E_{III,7}(\mathbf{w}\cdot\mathbf{G})\nabla_\perp T$$
$$+\left(E_{II}+E_{III,2}+E_{III,4}+E_{III,5}\right)(\mathbf{w}\cdot\mathbf{G})(\mathbf{G}\cdot\nabla_\perp T)\mathbf{G} + E_I(\mathbf{w}\cdot\mathbf{G})(\mathbf{G}\cdot\nabla_\perp N_e)\mathbf{G}, \tag{A.18}$$

where the factors are

$$E_I = -\frac{n_\kappa^{(e)}n_\kappa'^{(e)}A_\kappa A_\kappa'}{N_e}\left(\frac{m}{kT}\right)^2\frac{2}{\kappa+1}, \tag{A.19}$$

$$E_{II} = \frac{3n_\kappa^{(e)}n_\kappa'^{(e)}A_\kappa A_\kappa'}{2T}\left(\frac{m}{kT}\right)^2\frac{2}{\kappa+1}, \tag{A.20}$$

$$E_{III,1} = -n_\kappa^{(e)}n_\kappa'^{(e)}A_\kappa A_\kappa'\frac{m}{kT^2}\frac{2\kappa-3}{2\kappa+2}\left[1+\frac{1}{2\kappa-3}\frac{m}{kT}\left(G^2+\frac{w^2}{4}\right)\right], \tag{A.21}$$

$$E_{III,2} = n_\kappa^{(e)}n_\kappa'^{(e)}A_\kappa A_\kappa'\frac{m^2}{k^2T^3}\frac{1}{\kappa+1}, \tag{A.22}$$

$$E_{III,3} = n_\kappa^{(e)}n_\kappa'^{(e)}A_\kappa^2 A_\kappa'^2\frac{m^2}{2k^2T^3}\frac{(2\kappa-3)^2}{(2\kappa+2)^3}$$
$$\times\left\{G^2+\frac{w^2}{4}+\frac{1}{2\kappa-3}\frac{m}{kT}\left[\left(G^2+\frac{w^2}{4}\right)^2-(\mathbf{w}\cdot\mathbf{G})^2\right]\left[2+\frac{1}{2\kappa-3}\frac{m}{kT}\left(G^2+\frac{w^2}{4}\right)\right]\right\}, \tag{A.23}$$

$$E_{III,4} = n_\kappa^{(e)}n_\kappa'^{(e)}A_\kappa^2 A_\kappa'^2\frac{m^2}{k^2T^3}\frac{(2\kappa-3)^2}{(2\kappa+2)^3}\left\{1+\left(\frac{1}{2\kappa-3}\frac{m}{kT}\right)^2\left[\left(G^2+\frac{w^2}{4}\right)^2-(\mathbf{w}\cdot\mathbf{G})^2\right]\right\}, \tag{A.24}$$

$$E_{III,5} = n_\kappa^{(e)}n_\kappa'^{(e)}A_\kappa^2 A_\kappa'^2\frac{m^3}{k^3T^4}\frac{(2\kappa-3)}{(2\kappa+2)^2}\left\{2G^2+\frac{w^2}{2}+\frac{1}{2\kappa-3}\frac{m}{kT}\left[\left(G^2+\frac{w^2}{4}\right)^2-(\mathbf{w}\cdot\mathbf{G})^2\right]\right\}, \tag{A.25}$$

$$E_{III,6} = n_\kappa^{(e)}n_\kappa'^{(e)}A_\kappa^2 A_\kappa'^2\frac{m^3}{k^3T^4}\frac{(2\kappa-3)}{(2\kappa+2)^2}(\mathbf{w}\cdot\mathbf{G})^2, \tag{A.26}$$

$$E_{III,7} = -n_\kappa^{(e)} n_\kappa'^{(e)} A_\kappa A_\kappa' \frac{m}{kT^2} \frac{2\kappa-3}{2\kappa+2}. \tag{A.27}$$

In the above derivations, we have employed the relation,

$$\frac{w^2 \mathbf{U} - \mathbf{ww}}{w^3} \cdot \mathbf{w} = \frac{w^2 \mathbf{w} - \mathbf{w}(\mathbf{w}\cdot\mathbf{w})}{w^3} = 0. \tag{A.28}$$

Consider the integral with $(\mathbf{w}\cdot\mathbf{G})^2$ for an arbitrary even function $f(w,G)$,

$$\int_{-\infty}^{+\infty} f(w,G)(\mathbf{w}\cdot\mathbf{G})^2 \, \mathrm{d}\mathbf{w}\mathrm{d}\mathbf{G}. \tag{A.29}$$

The odd part of $(\mathbf{w}\cdot\mathbf{G})^2$ has no contribution to the above integral, so we have

$$\int_{-\infty}^{+\infty} f(w,G)(\mathbf{w}\cdot\mathbf{G})^2 \, \mathrm{d}\mathbf{w}\mathrm{d}\mathbf{G} = \int_{-\infty}^{+\infty} \frac{1}{3} f(w,G) w^2 G^2 \mathrm{d}\mathbf{w}\mathrm{d}\mathbf{G}. \tag{A.30}$$

In additional, it is easy to prove that $\mathrm{d}\mathbf{w}\mathrm{d}\mathbf{G} = \mathrm{d}\mathbf{v}\mathrm{d}\mathbf{v}'$ by calculating the Jacobian determinant. Using these expressions and substituting Eqs.(A.17)-(A.27) into Eq.(45), we derived the heat flux $\mathbf{q}_{ee}^{(e,2)}$,

$$\mathbf{q}_{ee}^{(e,2)} = \frac{2\pi e^4 L_e}{9 m \omega_{Be}^2} \int (\boldsymbol{\alpha}_1 + \boldsymbol{\alpha}_2 + \boldsymbol{\alpha}_3) \mathrm{d}\mathbf{w}\mathrm{d}\mathbf{G}, \tag{A.31}$$

with

$$\boldsymbol{\alpha}_1 = (E_{III,1} + E_{III,3} + E_{III,6} + E_{III,7}) w G^2 \nabla_\perp T, \tag{A.32}$$

$$\boldsymbol{\alpha}_2 = \frac{2}{3}(E_{II} + E_{III,2} + E_{III,4} + E_{III,5}) w G^4 \nabla_\perp T, \tag{A.33}$$

$$\boldsymbol{\alpha}_3 = \frac{2}{3} E_I w G^4 \nabla_\perp N_e. \tag{A.34}$$

In Eqs.(A.31)-(A.34), all the terms can be written as the two uniform mathematical form,

$$f_\kappa \int n_\kappa^{(e)} n_\kappa'^{(e)} A_\kappa A_\kappa' w^m G^n \mathrm{d}\mathbf{w}\mathrm{d}\mathbf{G}, \tag{A.35}$$

or

$$f_\kappa \int n_\kappa^{(e)} n_\kappa'^{(e)} A_\kappa^2 A_\kappa'^2 w^m G^n \mathrm{d}\mathbf{w}\mathrm{d}\mathbf{G}, \tag{A.36}$$

where $f_\kappa$ is an arbitrary factor which has no relation with integral, so that it can be taken out of the integral. These two types of integral are discussed in Appendix D. Substituting these results into Eqs. (A.31)-(A.34), we obtain the final expression,

$$\mathbf{q}_{ee}^{(e,2)} = -\frac{8}{3} \frac{\sqrt{\pi} e^4 L_e N_e^2 k^{1/2}}{m^{3/2} \omega_{Be}^2 T^{3/2}} C_\kappa \nabla_\perp T$$

$$+ \frac{2048}{27} \frac{g_\kappa \nu_{ei} e^4 L_e N_e^2 \sqrt{kT}}{\sqrt{2\pi} m^{3/2} \omega_{Be}^3 T} \left[\frac{\Gamma(\kappa+1)}{\Gamma(\kappa-1/2)}\right]^2 (\kappa+1)(2\kappa-3)^{1/2} H(4,1,\kappa+2) \mathbf{b} \times \nabla_\perp T, \tag{A.37}$$

where

$$C_\kappa = \frac{8\sqrt{2}}{3\sqrt{\pi}} \left[ \frac{\Gamma(\kappa+1)}{\Gamma(\kappa-1/2)} \right]^2 (\kappa+1)\left(\kappa-\frac{3}{2}\right)^{1/2} \left\{ \left[ 8 - \frac{9}{2(\kappa+2)} \right] H(2,1,\kappa+2) \right.$$

$$-\left[ 28 + \frac{25}{6(\kappa+2)} + \frac{44(\kappa+1)}{3(\kappa+2)} \right] H(4,1,\kappa+2) + \left[ 4 - \frac{23}{6(\kappa+2)} \right] H(2,3,\kappa+2)$$

$$-\frac{32}{9} H(6,3,\kappa+3) - \frac{40}{9} H(4,5,\kappa+3) - \frac{8}{3} H(4,1,\kappa+3) - \frac{8}{3} H(8,1,\kappa+3)$$

$$\left. -\frac{4}{3}(\kappa+1)\left[ 4H(6,1,\kappa+3) + 12H(4,3,\kappa+3) \right] \right\},$$

(A.38)

with the *H* function defined by

$$H(n,m,\kappa+2) = \int_0^\infty \left( 1 + 2x + 2y + \frac{2}{3}xy + x^2 + y^2 \right)^{-(\kappa+2)} x^{\frac{n+1}{2}} y^{\frac{m+1}{2}} dxdy.$$

**Appendix D:** Calculations of $\int n_\kappa^{(e)} n_\kappa^{\prime(e)} A_\kappa A_\kappa' w^m G^n d\mathbf{w} d\mathbf{G}$ and $\int n_\kappa^{(e)} n_\kappa^{\prime(e)} A_\kappa A_\kappa' w^m G^n d\mathbf{w} d\mathbf{G}$

$$\int n_\kappa^{(e)} n_\kappa^{\prime(e)} A_\kappa A_\kappa' w^m G^n d\mathbf{w} d\mathbf{G}$$

$$= N_e^2 B_\kappa^2 \left( \frac{2\kappa+2}{2\kappa-3} \right)^2 \int \left( 1 + \frac{1}{2\kappa-3} \frac{mv^2}{kT} \right)^{-(\kappa+2)} \left( 1 + \frac{1}{2\kappa-3} \frac{mv'^2}{kT} \right)^{-(\kappa+2)} w^m G^n d\mathbf{w} d\mathbf{G}$$

$$= N_e^2 B_\kappa^2 \left( \frac{2\kappa+2}{2\kappa-3} \right)^2 \int \left\{ 1 + \frac{1}{2\kappa-3} \frac{m}{kT} \left( 2G^2 + \frac{w^2}{2} \right) \right.$$

$$\left. + \left( \frac{1}{2\kappa-3} \frac{m}{kT} \right)^2 \left[ \left( G^2 + \frac{w^2}{4} \right)^2 - (\mathbf{w}\cdot\mathbf{G})^2 \right] \right\}^{-(\kappa+2)} w^m G^n d\mathbf{w} d\mathbf{G}.$$

We can expand the integrand into a series of $(\mathbf{w}\cdot\mathbf{G})^2$ by

$$\sum_n f_n(w,G)(\mathbf{w}\cdot\mathbf{G})^{2n},$$

(A.39)

where the factor $f_n(w,G)$ denotes the even function of $\mathbf{w}$ and $\mathbf{G}$. The odd part of $(\mathbf{w}\cdot\mathbf{G})^2$ has no contribution to the integral and therefore Eq.(A.39) can be rewritten as

$$\sum_n f_n(w,G) \left( \frac{w^2 G^2}{3} \right)^n.$$

(A.40)

Then the integrand is

$$\left\{ 1 + \frac{1}{2\kappa-3} \frac{m}{kT} \left( 2G^2 + \frac{w^2}{2} \right) + \left( \frac{1}{2\kappa-3} \frac{m}{kT} \right)^2 \left[ \left( G^2 + \frac{w^2}{4} \right)^2 - \frac{w^2 G^2}{3} \right] \right\}^{-(\kappa+2)} w^m G^n. \quad \text{(A.41)}$$

We let that

$$x = \left(\frac{1}{2\kappa-3}\frac{m}{kT}\right)G^2 \text{ and } y = \left(\frac{1}{2\kappa-3}\frac{m}{kT}\right)\frac{w^2}{4}, \quad (A.42)$$

so the integral can be rearranged as

$$\int n_\kappa^{(e)} n_\kappa'^{(e)} A_\kappa A_\kappa' w^m G^n \mathrm{d}\mathbf{w}\mathrm{d}\mathbf{G}$$

$$= N_e^2 \frac{16}{\pi}\left[\frac{\Gamma(\kappa+1)}{\Gamma(\kappa-1/2)}\right]^2 \left(\frac{2\kappa+2}{2\kappa-3}\right)^2 \left(\frac{1}{2\kappa-3}\frac{m}{kT}\right)^{-\frac{m+n}{2}} 2^{m+1} H(n,m,\kappa+2), \quad (A.43)$$

where the $H$ function is defined by

$$H(n,m,\kappa+2) = \int_0^\infty \left(1+2x+2y+\frac{2}{3}xy+x^2+y^2\right)^{-(\kappa+2)} x^{\frac{n+1}{2}} y^{\frac{m+1}{2}} \mathrm{d}x\mathrm{d}y. \quad (A.44)$$

Because of the similar calculation, the other integral is obtained as

$$\int n_\kappa^{(e)} n_\kappa'^{(e)} A_\kappa^2 A_\kappa'^2 w^m G^n \mathrm{d}\mathbf{w}\mathrm{d}\mathbf{G}$$

$$= N_e^2 \frac{16}{\pi}\left[\frac{\Gamma(\kappa+1)}{\Gamma(\kappa-1/2)}\right]^2 \left(\frac{2\kappa+2}{2\kappa-3}\right)^4 \left(\frac{1}{2\kappa-3}\frac{m}{kT}\right)^{-\frac{m+n}{2}} 2^{m+1} H(n,m,\kappa+3). \quad (A.45)$$

These two integrals are convergent under the condition $\kappa > 3/2$. By part integration, we can prove that the integral $H(n, m, \kappa+2)$ has a property which can be used in reduction of formulae, namely,

$$H(n,m,\kappa+2) = \int_0^\infty \left(1+2x+2y+\frac{2}{3}xy+x^2+y^2\right)^{-(\kappa+2)} x^{\frac{n+1}{2}} y^{\frac{m+1}{2}} \mathrm{d}x\mathrm{d}y$$

$$= \int \left(1+2x+2y+\frac{2}{3}xy+x^2+y^2\right)^{-(\kappa+2)} y^{\frac{m+1}{2}} \mathrm{d}\left(\frac{2}{n+3}\right) x^{\frac{n+3}{2}} \mathrm{d}y$$

$$= \frac{2(\kappa+2)}{n+3}\int \left(1+2x+2y+\frac{2}{3}xy+x^2+y^2\right)^{-(\kappa+2)} y^{\frac{m+1}{2}} x^{\frac{n+3}{2}} \left(2+2x+\frac{2}{3}y\right) \mathrm{d}x\mathrm{d}y$$

$$= \frac{2(\kappa+2)}{n+3}\left[H(n+2,m,\kappa+3)+H(n+4,m,\kappa+3)+\frac{2}{3}H(n+2,m+2,\kappa+3)\right], \quad (A.46)$$

where $n < 4\kappa+5$ is required. For the case of $\kappa > 3/2$, the condition turns to be $n < 11$, which is always satisfied in our clculations for all the terms. Noting that exchanging $n$ and $m$ has no effect on the integral, so one has its another property,

$$H(n,m,\kappa+2) = \frac{2(\kappa+2)}{m+3}\left[H(n,m+2,\kappa+3)+H(n,m+4,\kappa+3)+\frac{2}{3}H(n+2,m+2,\kappa+3)\right]. \quad (A.47)$$

**Appendix E:** The limit of $C_\kappa$ for $\kappa \to \infty$

First we calculate the term $H(n,m,\kappa+2)$ in the limit $\kappa \to \infty$. With the transformation (A.42), the integral can be rewritten as

$$\frac{1}{2^{m+1}}\left(\frac{1}{2\kappa-3}\frac{m}{kT}\right)^{\frac{m+n}{2}+3}\int\left\{\left(1+\frac{1}{2\kappa-3}\frac{mv^2}{kT}\right)\left(1+\frac{1}{2\kappa-3}\frac{mv'^2}{kT}\right)\right\}^{-(\kappa+2)}G^{n+2}w^{n+2}\mathrm{d}w\mathrm{d}G. \quad \text{(A.48)}$$

In the limit $\kappa \to \infty$, we have that

$$\lim_{\kappa\to\infty}\left\{\left(1+\frac{1}{2\kappa-3}\frac{mv^2}{kT}\right)\left(1+\frac{1}{2\kappa-3}\frac{mv'^2}{kT}\right)\right\}^{-(\kappa+2)} = e^{-\frac{m}{2kT}(v^2+v'^2)} = e^{-\frac{m}{kT}\left(\frac{w^2}{4}+G^2\right)}, \quad \text{(A.49)}$$

therefore $H(n,m,\kappa+2)$ tends to

$$\Gamma\left(\frac{n+3}{2}\right)\Gamma\left(\frac{m+3}{2}\right)(2\kappa-3)^{-\frac{m+n}{2}-3}. \quad \text{(A.50)}$$

Using the above relation and the limit of gamma function,

$$\lim_{n\to\infty}\frac{\Gamma(n+\alpha)}{\Gamma(n)n^{\alpha}}=1, \quad \text{(A.51)}$$

one finds that

$$\lim_{\kappa\to\infty}C_{\kappa}=1. \quad \text{(A.52)}$$